\documentclass[12pt,preprint]{aastex}
\usepackage{amssymb}

\shorttitle{Gamma-ray Burst Prompt Emission}

\shortauthors{Mao \& Wang}

\begin{document}


\title{Gamma-ray Burst Prompt Emission: Jitter Radiation in Stochastic Magnetic Field Revisited}


\author{Jirong Mao}
\affil{International Center for Astrophysics, Korea Astronomy and
Space Science Institute, 776, Daedeokdae-ro, Yuseong-gu, Daejeon
305-348, Republic of Korea} \affil{INAF-Osservatorio Astronomico
di Brera, Via Emilio. Bianchi, 46, 23807 Merate (LC), Italy}
\affil{Yunnan Observatory, Chinese Academy of Sciences, Kunming,
Yunnan Province, 650011, China} \affil{Key Laboratory for the
Structure and Evolution of Celestial Objects, Chinese Academy of
Sciences, Kunming, China}

\author{Jiancheng Wang}
\affil{Yunnan Observatory, Chinese Academy of Sciences, Kunming,
Yunnan Province, 650011, China} \affil{Key Laboratory for the
Structure and Evolution of Celestial Objects, Chinese Academy of
Sciences, Kunming, China}

\email{jirong.mao@brera.inaf.it}

\begin{abstract}
We revisit the radiation mechanism of relativistic electrons in
the stochastic magnetic field and apply it to the high-energy
emissions of gamma-ray bursts (GRBs). We confirm that jitter
radiation is a possible explanation for GRB prompt emission in the
condition of a large electron deflection angle. In the turbulent
scenario, the radiative spectral property of GRB prompt emission
is decided by the kinetic energy spectrum of turbulence. The
intensity of the random and small-scale magnetic field is
determined by the viscous scale of the turbulent eddy. The
microphysical parameters $\epsilon_e$ and $\epsilon_B$ can be
obtained. The acceleration and cooling timescales are estimated as
well. Due to particle acceleration in magnetized filamentary
turbulence, the maximum energy released from the relativistic
electrons can reach a value of about $10^{14}$ eV. The GeV GRBs
are possible sources of high-energy cosmic-ray.
\end{abstract}




\section{Introduction}
It has been widely discussed that the random and small-scale
magnetic field can be generated by Weibel instability
\citep{weibel59}. The initial magnetic field is formed from the
anisotropic distributed plasma disturbed by the certain
perturbation. This magnetic field can be amplified by induced
currents and finally reach the saturated value. This kind of
instability has been considered to occur in the relativistic
plasma \citep{yoon87,medvedev99,medvedev05}. The growth and
transport of this random and small-scale magnetic field have been
confirmed by numerical simulation
\citep{kazimura98,silva03,frederiksen04,hededal05} as well.

Jitter radiation, which is the emission of relativistic electrons
in this random and small-scale magnetic field, has been fully
investigated in recent years. This radiative mechanism, which is
different from the synchrotron process, has been applied
successfully to the research of some celestial objects, such as
gamma-ray bursts \citep{medvedev00,medvedev06}, their afterglows
\citep{medvedev07,workman08,morsony09}, and jets in the active
galaxies \citep{mao07}. Furthermore, \citet{medvedev09a} have
shown that anisotropy of the jitter radiation pattern and
relativistic kinetics can produce the time-resolved spectra of
prompt GRB emission. As the magnetic field may play a vital role
in the prompt emission of GRB jets
\citep{lyutikov01,giannios05,mckinney10}, another theory has also
been developed. The general theory of relativistic electron
radiation in a sub-Larmor scale magnetic field relevant to the
jitter regime includes the anisotropic magnetic field, the effects
of trapped electrons, and the case of a large deflection angle of
electrons \citep{medvedev10}.

The application of the jitter mechanism to GRB prompt emission has
been further analyzed. According to a simulation by
\citet{sironi09}, jitter radiation can be realized when the
strength of electromagnetic fields is reduced so that the wiggler
number $K=eBl_{\rm{cor}}/m_ec^2$ is smaller than unity, where $B$
is the magnetic field strength and $l_{\rm{cor}}$ is the length
scale of the magnetic field. \citet{kirk10} pointed out that the
standard shock scatterings are too weak to provide electrons with
the required Lorentz factor.

Considering the aforementioned, it is necessary to revisit the
jitter mechanism under the GRB physical conditions. In this paper,
we investigate the GRB prompt emission radiated by the
relativistic electrons in the stochastic magnetic field.
Turbulence, as the dominated physical point, is introduced to the
magnetic field generation and the particle acceleration within the
framework of the fireball model \citep{piran99}. Generally, the
GRB prompt spectrum can be fitted by the Band function
\citep{band93}. As the radiative spectral index below the
$E_{\rm{peak}}$ has been fully discussed by \citet{medvedev09a},
in this work we do not expect to fit the detailed spectrum for the
certain GRB source. Instead, we attempt to provide the general
radiative spectral property above $E_{\rm{peak}}$.

The possibility that GRB sources are linked with cosmic rays was
proposed by \citet{waxman95}, \citet{vietri95}, and
\citet{dermer04}. \citet{dermer01} described the stochastic
acceleration of GRB blast waves. The effects of heavy nuclei
accelerated by the GRB internal/external shocks have also been
taken into account \citep{murase08,wang08}. The relativistic
electrons radiated in the random magnetic field were considered by
some to solve the origin of high-energy cosmic rays
\citep{honda05,gureev08,honda09}. However, \citet{medvedev09b}
concluded that cosmic rays can be linked to the magnetic field
generated by Weibel instability as well. These results encouraged
us to further investigate the possibilities of cosmic-ray
production by jitter radiation under the GRB blast wave framework.

Here, we emphasize the energy of GRB prompt emission released from
relativistic electrons in the random magnetic field generated by
turbulence. We focus on three questions that are essential to the
radiative and turbulent processes: (1) Does the specific
simplified jitter radiation that can reproduce the spectral shape
and the energy of GRB prompt emission indeed exist; (2) Within the
framework of the GRB fireball model, how can the random and
small-scale magnetic field be generated by the turbulence, and
what is the dominated acceleration process for those relativistic
electrons; and (3) In our scenario, can GRBs be linked to
high-energy cosmic-ray sources?

In this paper, we extend our turbulent treatment of jitter
radiation \citep{mao07} to the investigation of GRB prompt
emission. In Section 2, after reviewing our specific jitter
radiation and the stochastic magnetic field, we apply this
radiative pattern to GRB prompt emission. We propose that the
spectral index of GRB prompt emission is determined by the energy
spectrum of turbulence. In Section 3, considering the stochastic
magnetic field is decided by the turbulent property, we attempt to
obtain the turbulent eddy scale under the GRB physical conditions.
To this end, some important parameters, such as $\epsilon_B$,
$\epsilon_e$, and electron Lorentz factor, can be obtained. In
order to further identify whether jitter radiation is valid, the
acceleration timescale, cooling timescale, and wiggler number are
estimated. We also speculate, within this specific jitter regime,
due to the turbulent effect, that GRBs are possible sources of
high-energy. Conclusions are given in Section 4.

\section{GRB Prompt Emission}
In this section, we propose that the random and small-scale
magnetic field is generated by turbulence. The relativistic
electron radiation in the random and small-scale magnetic field is
simplified in a one-dimensional case. In this specific jitter
regime, the GRB radiative spectral shape is determined by the
turbulent energy spectrum.

\subsection{Stochastic Magnetic Field}
In this work, we simply review the concept of the stochastic
magnetic field generated by turbulence \citep{mao07}. The energy
spectrum in a general turbulent field can be described as
\begin{equation}
F(q)\propto q^{-\zeta_p}.
\end{equation}
In the turbulent fluid, through the cascade process, the turbulent
energy dissipation field has a hierarchical fluctuation structure.
A set of inertial-range scaling laws of fully developed turbulence
can be derived. From the research of \citet{she94} and
\citet{she95}, the energy spectrum index $\zeta_p$ of the
turbulent field is related to the cascade process number $p$ by
the universal relation $\zeta_p=p/9+2[1-(2/3)^{p/3}]$. The
Kolmogorov turbulence is presented as $\zeta_p=p/3$. The typical
Kolmogorov numbers are $p=5$ and $\zeta_p=5/3$.

The stochastic magnetic field $<\delta B(q)>$ generated by the
turbulent cascade in one-dimension can be given by
\begin{equation}
<\delta B^2(q)>\sim K(q)\sim \int_q ^{\infty} F(q')dq',
\end{equation}
where $q_\nu<q<q_\eta$, $q_\nu$ is linked to the viscous
dissipation while $q_\eta$ is related to the magnetic resistive
transfer. The Prandtl number ${\rm{Pr}}=10^{-5}T^4/n$ constrains
the number of $q$ by $q_\eta/q_\nu={\rm{Pr}}^{1/2}$, where $T$ is
the temperature of the plasma and $n$ is the plasma number density
\citep{schekochihin07}.

\subsection{Jitter Radiation}
The radiation by a single relativistic electron in the small-scale
magnetic field was studied by \cite{landau71}. The radiation
intensity which is the energy per unit frequency per unit time is
\begin{equation}
I_\omega=\frac{e^2\omega}{2\pi
c^3}\int^{\infty}_{\omega/2\gamma_\ast^2}\frac{|\bf{w}_{\omega'}|^2}{\omega'^2}{(1-\frac{\omega}{\omega'\gamma_\ast^2}+\frac{\omega^2}{2\omega'^2\gamma_\ast^4})}d\omega'
,
\end{equation}
where $\gamma_\ast^{-2}=(\gamma^{-2}+\omega^2_{pe}/\omega^2)$,
$\omega'=(\omega/2)(\gamma^{-2}+\theta^2+\omega^2_{pe}/\omega^2)$
is the frequency in the radiative field, $\omega_{pe}=(4\pi
e^2n/m_e)^{1/2}$ is the background plasma frequency, and $\gamma$
is the electron Lorentz factor. This equation presents a general
description of diffusive synchrotron radiation. In principle, this
radiation is shown in three dimensions.

Following one specific treatment by \citet{mao07}, we simplify the
above equation in our one-dimensional case as
\begin{equation}
I_\omega=\frac{e^4}{m^2c^3\gamma^2}\int^{\infty}_{1/2\gamma_\ast^2}d(\frac{\omega'}{\omega})(\frac{\omega}{\omega'})^2{(1-\frac{\omega}{\omega'\gamma_\ast^2}+\frac{\omega^2}{2\omega'^2\gamma_\ast^4})}
\int{dq_0dq\delta(w'-q_0+qv)K(q)\delta[q_0-q_0(q)]}.
\end{equation}
Some interesting cases of the radiative spectrum were fully
discussed by \citet{medvedev00,medvedev06}. In this context, we
note that the radiation field is strongly associated with the
perturbation field. Thus, the adopted dispersion relation
$q_0=q_0(q)$ is important for the radiative property. In
particular, at high frequency, Equation (4) can be simplified by
the certain dispersion relation, and the radiative property is
determined by the structure of the magnetic field \citep{mao07}.

\subsection{Spectrum of GRB Prompt Emission}
We first calculate the relativistic electron frequency
$\omega_{pe}=(4\pi e^2n/\Gamma_{\rm{sh}}m_e)^{1/2}=9.8\times
10^9\Gamma_{\rm{sh}}^{-1/2}~\rm{s^{-1}}$, where $\Gamma_{\rm{sh}}$
is the bulk Lorentz factor of the shock. Here, we take the value
$n=3\times 10^{10}~\rm{cm^{-3}}$ as the number density in the
relativistic shock; this is consistent with the value used by
\citet{medvedev99}. From Equation (4), we see that the radiative
frequency is associated with the dispersion relation. Here, we
refer to the work of \citet{milosavljevic06}. In their research,
the steady state from Weibel instability in the relativistic shock
was fully discussed. Having the view angle $\theta\sim
\gamma^{-1}$ in the radiative field, we solve that dispersion
relation and get
\begin{equation}
\omega=\gamma^2cq[(1\pm \sqrt{1-4\omega^2_{pe}/\gamma
c^2q^2})/2]^{1/2}.
\end{equation}
Assuming $\gamma c^2q^2\gg 4\omega^2_{pe}$, we obtain $\omega\sim
\gamma^2cq$.

Then, we can integrate Equation (4). At the high-energy band, the
radiation frequency $\omega\gg \omega_{pe}$. After inserting the
dispersion relation (5) and Equation (2) in Equation (4), we
obtain
\begin{equation}
I_\omega\sim \omega^{-(\zeta_p -1)}.
\end{equation}
As the single electron spectral index is not affected by the
electron Lorentz factor $\gamma$, this spectral shape can be
viewed as the final result of gross radiative plasma.

However, in our specific case, we further note, as $\zeta_p$ is
the factor of turbulent scaling law, that the spectral shape of
GRB high-energy radiation is determined by the microphysical
turbulent property. Thus, we determine one physical reason to
interpret this power-law radiative spectrum shown in Equation (6):
The radiative spectral index is fully decided by the cascade
process of the turbulent field.

The typical value of $\zeta_p$ is the Kolmogorov number 5/3, given
the cascade $p=5$. However, this number is not universal. As shown
in Section 2.1, the different value of $\zeta_p$ corresponds to
the different cascade number $p$ \citep{she94,she95}. Therefore,
from Equation (6), the radiative spectral index of GRB high-energy
emission is allowed a wide range. Moreover, in this paper, we
consider the simulation results given by \citet{schekochihin04}.
From their research, the physical processes on the small-scale
turbulent dynamo were extensively illustrated. To our interests,
in the condition of large Reynolds numbers, the turbulent fluid is
viscous dominated and the Prandtl number ${\rm{Pr}}\gg 1$. After
the short segment of Kolmogorov scaling index 5/3, the saturated
energy spectrum is deeper and the index turns to the value $7/3$.
This simulation result corresponds to the cascade number $p=7$;
thus, we have $\zeta_p=7/3$ for Kolmogorov form and $\zeta_p=2.0$
for the calculation from \citet{she94} and \citet{she95}.

Finally, we can compare our investigations to the GRB prompt
emission data observed by high-energy satellites. From the Burst
And Transient Source Experiment observation, above
$E_{\rm{peak}}$, the flux $f_\nu$ has a power-law index of
$1.25$--$1.4$ \citep{preece00,preece02}. From the rough statistics
of {\it{Fermi}} Gamma-ray Burst Monitor (GBM) data by
\citet{guetta09}, \citet{ghisellini10b} and \citet{bissaldi11},
the spectral index above $E_{\rm{peak}}$ has a similar value.
Thus, the case in which the radiative spectral index $4/3$ is
derived from the cascade number $p=7$ and the turbulent energy
spectrum index $\zeta_p=7/3$ from Kolmogorov form may be one
possibility that fits the above observational results. For the
following calculations, we take $\zeta_p=7/3$ as the reference
number.

\section{Magnetic Field and Particle Acceleration of GRBs}
Due to the turbulent effect, the magnetic field can be derived by
the scale of the viscous eddy. The electrons can be accelerated to
the maximum Lorentz factor about $10^{12}$ by the scattering in
the magnetized filaments. In this scenario, we identify that
jitter radiation is valid for the GRB prompt emission and GRBs
with maximum energy of about $10^{14}$ eV might be linked to the
possible cosmic-ray sources.

\subsection{Magnetic Field}
From Equation (2), the magnetic field can be estimated by the
following calculation:
\begin{equation}
<B>=[\int_{q_{\nu}}^{q_\eta} q^{-\zeta_p}dq]^{1/2} =
q_{\nu}^{(1-\zeta_p)/2}/\sqrt{\zeta_p-1}
\end{equation}
with the condition $q_\eta\gg q_\nu$. Since $q_\nu$ is identified
by the viscous property, we take the scale of the viscous eddy to
quantify it \citep{kumar09,narayan09,lazar09}. Thus, the number
$q_\nu$ can be calculated as
\begin{equation}
q_\nu=2\pi
l^{-1}_{\rm{eddy}}=2\pi(R/\Gamma_{\rm{sh}}\gamma_t)^{-1}=6.3\times
10^{-10}(\frac{R}{10^{13}~\rm{cm}})^{-1}(\frac{\Gamma_{\rm{sh}}}{100})(\frac{\gamma_t}{10})~\rm{cm^{-1}},
\end{equation}
where $\Gamma_{\rm{sh}}$ and $\gamma_t$ are the Lorentz factor of
shock and the turbulent eddy, respectively. At the fireball radius
$10^{13}$ cm, taking the turbulent spectrum $\zeta_p=7/3$, we have
a magnetic field of about $1.2\times 10^6$ G. This result is fully
consistent with the estimation from the fireball internal shock
model \citep{piran05}. As presented in Section 2.1, the value
$q_\eta$ can be obtained by $q_\eta=Pr^{1/2}q_\nu$. We can obtain
the plasma temperature by the form $T=m_ec^2/k$; $k$ is the
Boltzmann constant. Finally, we have magnetic resistive scale
$q_\eta$ as
\begin{equation}
q_\eta=3.9\times 10^2(\frac{n}{3\times 10^{10}
~\rm{cm^{-3}}})^{-1/2}(\frac{T}{5.6\times
10^9~\rm{K}})^2~\rm{cm^{-1}}.
\end{equation}
Moreover, the micro-parameter $\epsilon_B$ can be achieved as
\begin{equation}
\epsilon_B=5.2\times 10^{-2}(\frac{B}{1.2\times 10^6
~\rm{G}})^2(\frac{\Gamma}{100})^3(\frac{\delta
t}{1~\rm{s}})^3(\frac{E_k}{10^{51}~\rm{erg}})^{-1},
\end{equation}
where $\delta t$ is the time variability of the radiative pulse
and $E_k$ is the kinetic energy of the relativistic shell.
Following the estimation from \citet{medvedev06b} we can obtain
$\epsilon_e=\sqrt{\epsilon_B}=0.23$ through the typical value
$\epsilon_B=0.052$.

\subsection{Acceleration and Timescales}
Particle acceleration is strongly involved in the physical
processes described in Section 3.1. \citet{hededal04} found that
the relativistic electrons have a power-law distribution, but the
acceleration is local. \citet{nishikawa06} confirmed that the
electrons are not {\it{Fermi}}-accelerated and the processes in
the relativistic collisionless shock are dominated by Weibel
instability. The three-dimensional simulation presented in detail
that the Weibel instability excited in collisionless shocks is
responsible for electron acceleration \citep{nishikawa09}. In this
paper, we demonstrate that, under our scenario, electron
acceleration in the magnetized current filaments is vital for the
jitter radiation.

Electron energy distribution is not a power law under a turbulent
process. \citet{schli84,schli85} already found an
ultra-relativistic Maxwellian energy distribution.
\citet{stawarz08} used this Maxwellian energy distribution to
produce the synchrotron and inverse Compton spectra. Recently, the
Maxwellian electron component has been found through simulation
\citep{spitkovsky08} and has been applied to GRB research
\citep{giannios09}. In this context, we estimate the average
Lorentz factor of relativistic electrons given by
\citet{giannios09}: $<\gamma>=\epsilon_e\Gamma m_p/m_e$. Using the
number $\epsilon_e=0.23$, which we have derived, and $\Gamma=100$,
we obtain $<\gamma>=3.6\times 10^4$.

In general, stochastic acceleration has been fully discussed by
\citet{schli89a,schli89b}, \citet{dermer96}, and
\citet{stawarz08}. The acceleration timescale was calculated by
\citet{schli89a,schli89b} and \citet{petrosian04}. From their
studies, the particles are buried in a regular and external
magnetic field. The magnetic field is generated by macroscopic
turbulence \citep{sironi07}. However, this is not our physical
situation. Alternatively, \citet{honda05} and \citet{honda09}
mentioned the acceleration process in the random magnetic field.
The electrons can be effectively accelerated in the magnetized
current filaments. This acceleration process is consistent with
the physics of Weibel instability and our magnetic field
generation. Here, we apply this kind of process. The acceleration
timescale can be calculated as
\begin{equation}
t_{\rm{acc}}=(\frac{6^{1/2}\pi}{8})(\frac{c}{L})(\frac{E}{eBU})^2
=1.4\times 10^{-12}(\frac{E}{\rm{MeV}})^2(\frac{B}{1.2\times
10^6~\rm{G}})^{-2}(\frac{L}{1.0\times
10^{10}~\rm{cm}})^{-1}({\frac{U}{0.1c}})^{-2}~\rm{s},
\end{equation}
where we take the upstream speed $U\sim 0.1c$. The turbulent
length scale $L$ should be given by $q_\eta < L^{-1} < q_\nu$. In
the equation above, we use $L\sim 10^{10}$ cm as a reference
number. The cooling timescale can be estimated as
\begin{equation}
t_{\rm{cool}}=\frac{6\pi m_ec}{\sigma_T\gamma B^2} =1.5\times
10^{-8}(\frac{\gamma}{3.6\times 10^4})^{-1}(\frac{B}{1.2\times
10^6~\rm{G}})^{-2}~\rm{s}.
\end{equation}
From the calculations above, at the region $10^{13}$ cm from the
burst center, for a single electron with the average Lorentz
factor $\gamma\sim 10^4$, at the MeV band, we have
$t_{\rm{acc}}<t_{\rm{cool}}$, meaning that the particle
acceleration is effective for jitter radiation.

Recently, \citet{medvedev09c} investigated the cooling timescale
in detail. The relativistic plasma frequency of the protons is
given by $\omega_{pp}=3.8\times 10^8(n/3\times
10^{10}\rm{cm^{-3}})^{1/2}~\rm{s^{-1}}$; we can obtain the
parameter $T_{\rm{cool}}=t_{\rm{cool}}\omega_{pp}<(10-50)$.
Following their analysis, this means that cooling is very fast
with strong radiative loss; thus, {\it{Fermi}} acceleration is
impossible. In our scenario, the electron scattering in the
magnetized filaments can be a self-consistent way for the particle
acceleration.

In order to further illustrate the validation of jitter radiation,
first, we note that the characteristic correlation scale of the
magnetic field $l_{\rm{cor}}$ should be less than the typical
Larmor radius $r_L$ of a relativistic electron. The correlation
scale $l_{\rm{cor}}\sim (0.1-1)l_{\rm{sk}}$,
$l_{\rm{sk}}=c/\omega_{pe}$ is the skin depth. For a relativistic
electron the Larmor radius is $r_L\sim \gamma m_ec^2/eB$. If
$l_{\rm{cor}}<r_L$, we obtain $\gamma>2.2\times 10^4$ for a
magnetic field $B\sim 1.2\times 10^6~\rm{G}$ and $l_{\rm{cor}}\sim
l_{sk}$. Then, we calculate the wiggler parameter
\citep{sironi09,medvedev10} as
\begin{equation}
K=\frac{eBl_{\rm{cor}}}{m_ec^2}=2.2\times 10^4(\frac{B}{1.2\times
10^6~\rm{G}})(\frac{\Gamma_{\rm{sh}}}{100})^{1/2}(\frac{n}{3\times
10^{10}~\rm{cm^{-3}}})^{-1/2}.
\end{equation}
Thus, we obtain $1<K<\gamma$. Following the analysis by Medvedev
et al. (2010), the jitter regime can be satisfied in the condition
of a large electron deflection angle. We further calculate the
deflection angle $\theta=eBl_{\rm{cor}}/\gamma m_ec^2$. using
numbers $B\sim 1.2\times 10^6~\rm{G}$ and $\Gamma_{\rm{sh}}\sim
100$, we obtain $\theta < 1$.

\subsection{GeV GRBs and Possible Cosmic-ray Origin}
We further obtain some interesting results of high-energy emission
associated with the turbulence. As mentioned by \citet{honda05},
the particles can be accelerated by interaction with the local
magnetic filaments. \citet{virtanen05} studied the electron energy
gained by turbulent scattering. Since we have obtained the
magnetic field by the turbulent viscous eddy, we can calculate the
maximum Lorentz factor of electrons as
\begin{equation}
\gamma_{e,{\rm{max}}}=eB/q_\nu m_ec^2=1.1\times 10^{12}
\end{equation}
with $B=1.2\times 10^6~\rm{G}$. Thus, at distance $R\sim 10^{13}$
cm from the GRB explosion center, due to relativistic turbulence,
the acceleration process can produce electrons with extremely high
Lorenz factor values up to $10^{12}$. This value is much larger
than the estimation number of \citet{kirk10}. The lower Lorentz
factor of electrons given by \citet{kirk10} comes from
{\it{Fermi}} acceleration \citep{achterberg01}.

As the turbulent region is $10^{13}$--$10^{16}$ cm from the burst
center, where $R\sim 10^{13}$ cm is the optical-thin radius and
$R\sim 10^{16}$ cm is the deceleration radius of the fireball, we
obtain the maximum frequency of the jitter radiation
$\omega_{\rm{max}}\sim 10^{11}-10^{18}$ eV. The maximum value is
strongly dependent on the emission region related to the fireball
radius. In order to further constrain the released energy in the
jitter regime, we have the constraint that comes from the results
of Medvedev et al. (2010). In the jitter radiation with the
condition of a large electron deflection angle, the break
radiative frequency has been derived as $\omega_b\sim
(c/l_{\rm{cor}})^3(eB/\gamma m_ec)^{-2}$. Our calculations in this
paper are valid if we have the radiation frequency below this
break. As this break frequency dependent on $\gamma$, we obtain
the maximum value of the break $\omega_{b,{\rm{max}}}\sim
10^{11}~\rm{eV}$ if $B\sim 1.2\times10^6~\rm{G}$ and
$\Gamma_{\rm{sh}}\sim 100$. Finally, we use
$t_{\rm{acc}}=t_{\rm{cool}}$ and obtain the maximum possible
energy of electrons
\begin{equation}
E=2.5\times 10^{14}
(\frac{U}{0.1c})(\frac{L}{1.0\times10^{10}~\rm{cm}})^{1/2}(\frac{\gamma}{3.6\times
10^4})^{-1/2}~\rm{eV}.
\end{equation}
This maximum energy is dependent on the Lorenz factor $\gamma$ of
electrons and the turbulent length scale $L$.

Although some other mechanisms, such as thermal emission,
synchrotron self-Compton, and Comptonization, are proposed to
modify the spectrum shape \citep{toma10,peer10}, we see that the
MeV--GeV GRB emissions might be reproduced by jitter radiation in
our specific case. Some GRBs detected by {\it{Fermi}} satellite
may support our speculation. GRB 080916C \citep{abdo09} and GRB
090217A \citep{acke10} are the typical examples. In their spectra,
we cannot find any cutoff evidence toward the high-energy
frequency. With the same spectral index, the spectra might be
extrapolated to the higher energy band, which is out of the energy
threshold of {\it{Fermi}} observation. Thus, in our specific case,
the GRB jitter regime naturally provides the link to the
high-energy cosmic rays. These GRBs are the possible sources of
cosmic-ray.

The estimated electron maximum energy is roughly consistent with
the first knee at about $10^{15}~\rm{eV}$ seen in the cosmic-ray
spectrum \citep{antoni05}. Within the framework of our simple
treatment, this maximum energy is still below the
Greisen--Zatsepin--Kuzmin limit. We propose that the first knee at
about $10^{15}$ eV in the cosmic-ray spectrum might be due to
variations in electron Lorentz factor and related timescales in
our calculations. The energy observed above $10^{18}$ eV could be
explained by some other mechanisms in which the heavy nuclei are
involved \citep{honda09}.

Although our estimated maximum energy can reach the GeV band, in
general, above 100 MeV, with the average number $\gamma\sim 10^4$,
the acceleration timescale is larger than the cooling timescale,
and the jitter radiation does not work. Thus, in our scenario we
do not expect many more GeV GRBs.\footnote{See \citet{fan09} for
the "traditional" interpretations about non-detection GeV
emissions.}

The high-energy spectral properties are more complicated. With the
collection data from \citet{ghisellini10b}, about 11 GRBs observed
by {\it{Fermi}} Large Area Telescope (LAT), and the constraints by
\citet{guetta10}, we find that some GRB GeV emissions detected by
LAT can not be explained by our simple treatment. Recently,
\citet{kumar09b} and \citet{kumar10} proposed that GeV emission
detected by {\it{Fermi}} originates from the afterglow and is
produced by the external shock. This is questioned by
\citet{piran10}. From the calculation in the external shock model
\citep{barniol10}, the magnetic field is about $10~\rm{\mu G}$,
which is a very low number compared with our estimation. Thus, our
treatment cannot apply to the afterglow/external shock scenario.
Of course, other explanations for the GRB MeV-GeV emissions cannot
be ruled out \citep{zhang09,gao09,wang09,li10}.

\section{Conclusion}
Although jitter radiation as an explanation of GRB prompt emission
has recently been challenged by \citet{sironi09} and
\citet{kirk10}, by applying the specific case of relativistic
electron radiation in the random and small-scale magnetic field,
we have successfully reproduced the radiative spectral index above
energy $E_{\rm{peak}}$. The spectral variability of GRB emission
below $E_{\rm{peak}}$ has been determined by \citet{medvedev09a},
so the problem of the "synchrotron line of the death"
\citep{preece98,savchenko09} might be solved as well.

In our scenario, the random and small-scale magnetic field is
generated by turbulence. The GRB radiative property in the jitter
regime is determined by the turbulent structure. We also declare
that particle acceleration under Weibel instability in the
relativistic shock may differ from {\it{Fermi}} acceleration. As
{\it{Fermi}} acceleration is inefficient for jitter radiation, in
our work, the acceleration process in the magnetized filamentary
turbulence plays a key role in explaining the GRB prompt
emissions.

We have shown that jitter radiation can produce GRB high-energy
photons at the MeV-GeV band. We further suggest that our specific
jitter regime of GRBs could be one of the explanations for
high-energy cosmic-ray origin. In contrast, \citet{zou09} have
discussed many kinds of inverse Compton processes for GRB
high-energy emission. As pointed out by \citet{kirk10}, the
Compton mechanism, which produces photons from jitter radiation
scattered by the thermal/nonthermal electrons, may modify the
original spectral property. We will investigate this more
complicated process in future work.

\acknowledgments

We thank the referee for the constructive suggestions. J. Mao is
grateful to the researchers in INAF-OAB, Merate. We acknowledge
the financial support from the National Natural Science Foundation
of China 10778702, the National Basic Research Program of China
(973 Program 2009CB824800), and the Policy Research Program of the
Chinese Academy of Sciences (KJCX2-YW-T24).

\clearpage

\end{document}